 \documentclass[structabstract]{aa}
\usepackage{graphicx}
\usepackage{txfonts}
\usepackage{natbib}
\bibpunct{(}{)}{;}{a}{}{,} % to follow the~A&A style
\begin{document}
\title{ Really focused stellar winds in X-ray binaries}
% \subtitle{ }
\author{P. Hadrava\inst{1} \and  J. \v{C}echura\inst{1,2} %\inst{2}
% \fnmsep\thanks{This study uses
%  the~spectra from Ond\v{r}ejov 2-m telescope.}
 }
\institute{Astronomical~Institute, Academy~of~Sciences,
      Bo\v{c}n\'{\i}~II~1401, CZ~-~141~31~Praha~4,
      Czech~Republic\\
      \email{had@sunstel.asu.cas.cz}
       \and
 Faculty of Mathematics and Physics, Charles University, Prague,
      Czech~Republic\\
      \email{cechura@astro.cas.cz}
%     \thanks university of heaven temporarily does not accept e-mails}
             }

   \date{Received Nov 2, 2010; revised Feb 21, 2011} % AA/2010/16046}
% \abstract{}{}{}{}{}
% 5 {} token are mandatory
  \abstract
  {}
  {We investigate the~anisotropy of stellar winds in binaries to improve
  the~models of accretion in high-mass X-ray binaries.}
  {We model numerically the~stellar wind from a~supergiant component
 of a~binary in radial and three-dimensional radiation hydrodynamic 
 approximation taking into account the~Roche potential, Coriolis force, 
 and radiative pressure in the~continuum and spectral lines.}
  {The~Coriolis force influences substantially the~mass loss and thus 
 also the~accretion rate. The focusing of the~stellar wind by
 the~gravitational field of the~compact companion leads to the~formation
 of a~gaseous tail behind the~companion.}
  {}


   \maketitle
%
%________________________________________________________________
%\documentclass[a0,portrait,posthelv,seminar]{a0poster}
%\pagestyle{empty}
%\setcounter{secnumdepth}{0}
%\usepackage[absolute]{textpos}
%\usepackage{wrapfig}
%\usepackage{times}
%\usepackage{eepic}
%\usepackage{pstricks}
%\usepackage[tiling]{pst-fill}
%\renewcommand{\cite}{\citeANP}
%\renewcommand{\VeryHuge}{\fontsize{89.16pt}{90pt}\selectfont}
%\newcommand{\KOREL}{{\sl KOREL}}
%\newcommand{\Halpha}{\ensuremath{{\mathrm H}\alpha}}
%\newcommand{\CCD}{{\em CCD}}
% These colours are tried and tested for titles and headers. Don't
% over use color!
%\usepackage{color}
%\definecolor{DarkBlue}{rgb}{0.1,0.1,0.5}
%\definecolor{Red}{rgb}{0.9,0.0,0.1}
%\definecolor{Blue}{rgb}{0.1,0.1,0.8}
%\definecolor{zelinkava}{rgb}{0.,0.75,0.43}
%\definecolor{Pink}{rgb}{0.5,0.1,0.1}
%
%\def\refname{\color{napis} References}
%\pagecolor{podklad}
%-------------------------------------------------------------------

\section{Introduction}

Roche-lobe overflow and an~isotropic radial stellar wind have traditionally
been considered as two competing scenarios to explain the~mass loss from 
a~companion star that can feed an~accretion disc around a~compact object,
particularly in interacting binaries and X-ray binaries -- cf., e.g. 
\cite{1973A&A....24..337S}. 
One of us thus developed a~model of an~evaporative radial stellar wind 
modulated by the~effective Roche potential to overcome this dichotomy 
\citep{HPhD}.  
This model revealed that there is a~smooth transition between the~outflow 
from a~narrow stream around the~inner Lagrangian point $L_{1}$ for small 
ratios of thermal to escape velocities, to an almost isotropic wind for 
high values of this ratio. 
The~distribution of mass-loss rates on the~surface of the~companion is 
modulated depending also on the~mass ratio of the~system. In some cases, 
the~mass loss in the~vicinity of the~$L_{2}$ point exceeds that around 
$L_{1}$. 
S. K\v{r}\'{\i}\v{z} (1980, private communication) raised an objection to
this model based on the~argument that it neglects the~Coriolis force and 
thus artificially enhances the~outflow by an~unrealistic 
centrifugal force. Because of this criticism, the~model was not published 
until \citet[cf. the~on-line Appendix]{1987PAICz..70..263H}.  
In the~meantime, a~similar model of a~radiatively driven wind was 
published by \cite{1982ApJ...261..293F}. 
This model of a~so-called focused stellar wind is now widely used to 
explain the~behaviour of X-ray binaries. However, the~above-mentioned 
objection is also valid for this model and the~Coriolis force should be 
taken into account to improve the~model of stellar wind anisotropy. This 
may significantly influence the~interplay between the~mass loss from 
the~donor star, its accretion by the~compact component, and the~consequent 
ionization of the~donor's wind and atmosphere illuminated by the~X-ray 
radiation -- cf. \cite{2008ApJ...678.1237G}. 

   Here we present our first preliminary results of our revised calculations 
of the~anisotropic stellar wind in binaries achieved by \cite{CDipl}.
In Section~\ref{radwind}, we summarize our calculations of the~radial wind 
based on the~Friend-Castor approximation. To reveal the~role of the~Coriolis 
force, we then numerically modelled on a~three-dimensional (3D) grid 
the~radiation hydrodynamics of the~stellar wind taking into account 
the~same expressions for the~radiative pressure in the~lines -- cf. 
Section~\ref{hydro}. These 
calculations showed the~formation of a~gaseous tail behind the~compact
companion, which resembles Bondi-Hoyle-Lyttleton accretion 
\citep{1939PCPS...34..405H} 
-- cf. Section~\ref{BHL}. As discussed in Section~\ref{dis}, this feature 
actually deserves to be denoted as a~gravitational focusing of the~wind 
rather than the~slight tidal modulation of the~wind's root.

\section{Radial stellar wind}\label{radwind}
Our present model of anisotropic radial stellar wind is based on a~modified 
version of the~line-driven wind theory of Castor, Abbott, and Klein 
\citep[CAK hereafter]{1975ApJ...195..157C}, 
where the~wind material is predominantly accelerated by the~line absorption 
of the~star's radiation field. In this work, the~authors solved the~equation 
of radiative transfer in the~stellar wind under the~assumption of 
the~Sobolev approximation \citep{1974MNRAS.169..279C} 
and found that the~line force could be approximated by a~power law in 
optical depth 
\begin{equation}\label{MOD1}
 f_{L} = \frac{\sigma_\mathrm{e}L_\ast}{4\pi{}cr^2}kt^{-\alpha}\; ,
\end{equation}
where \(t\) is the~optical depth parameter defined by
\begin{equation}\label{MOD2}
 t = \sigma_\mathrm{e}\rho{}\varv_{\mathrm{th}}\left(\frac{\mathrm{d}\varv}{\mathrm{d}r}\right)^{-1}\; ,
\end{equation}
and \(k\) and \(\alpha\) are parameters of the~CAK model that depend on 
the~effective temperature of the~star. Physically speaking, \(k\) is a~measure 
of the~number of driving lines, and \(\alpha\) is a measure of the~relative
distribution of optically thick to optically thin lines.  We chose 
the~values of \(\alpha=0.7\) and \(k=1/30\) 
in our model. These values should provide a reasonable approximation of 
the~line force corresponding to a~star with an~effective temperature of 
around 30000 K (cf. CAK). The~quantity \(\sigma_e\) is the~electron 
scattering coefficient, \(L_\ast\) is the~luminosity of the~primary star, 
\(\rho\) is the~density of the~wind material, and \(\varv_\mathrm{th}\) 
is the~thermal velocity of the~ions in the~wind.

 In the~case of an~isothermal radial outflow, the~equation of momentum
conservation is given by
\begin{eqnarray}\label{MOD3}
 \left(\varv-\frac{a^2}{\varv}\right)\frac{\mathrm{d}\varv}{\mathrm{d}r}= &-& \frac{\mathrm{d}\Phi_\mathrm{eff}}{\mathrm{d}r}+
 \frac{2a^2}{r}\ \nonumber\\* &+&\frac{\Gamma_\ast{}GM_\ast{}k}{r^2}\left(\sigma_\mathrm{e}\varv_\mathrm{th}\frac{\mathrm{d}\dot{M}}
 {\mathrm{d}\Omega}\right)^{-\alpha}\left(r^2\varv\frac{\mathrm{d}\varv}{\mathrm{d}r}\right)^{\alpha} ,
\end{eqnarray}
where \(a=\left(\mathrm{d}P_g/\mathrm{d}\rho\right)^{1/2}\) is the~isothermal 
sound speed, \(M_\ast\) is the~mass of the~primary star, and \(\Gamma_\ast\) 
is the~ratio of the~primary star luminosity to the~Eddington luminosity
\begin{equation}\label{MOD4}
 \Gamma_\ast = \frac{\sigma_\mathrm{e}L_\ast}{4\pi{}GM_\ast{}c}\; .
\end{equation}

   To include the~effects of the~compact companion on the~wind dynamics,
we have replaced the~spherically symmetric gravitational potential of 
the~mass-losing star by the~Roche potential \(\Phi_\mathrm{eff}\), i.e.
both the~tidal force and the~centrifugal force due to a~rotation synchronized 
with the~circular orbital motion of the~companion are taken into account. 
In a~spherical coordinate system, a new effective potential is given {by 
\begin{eqnarray}\label{MOD5}
  \Phi_\mathrm{eff}\left(r,\phi,\theta\right) = &-&\frac{GM_\ast}{D}\left\{\frac{D\left(1-\Gamma_\ast\right)}{r}\right. \nonumber \\ 
&+&\frac{q\left(1-\Gamma_\mathrm{x}\right)}{\left[1-2\left(r/D\right)\lambda+\left(r/D\right)^2\right]^{1/2}} \nonumber \\ &-&\left.q\left(\frac{r}{D}\right)\lambda + \frac{1}{2}\left(1+q\right)\left(\frac{r}{D}\right)^2\left(1-\mu^2\right)\right\}\; ,
 \end{eqnarray}
where the~variables \(\lambda\) and \(\mu\) are defined by 
\begin{eqnarray}\label{MOD5b}
  \lambda &=& \cos\phi\cos\theta\; , \nonumber \\ 
  \mu     &=& \sin\phi \; ,
 \end{eqnarray}
in terms of the~latitude \(\phi\) and the~longitude \(\theta\) on the~surface
of the~mass-losing star, \(D\) is the~separation of the~components of 
the~binary, \(q\) is the~ratio \(M_\mathrm{x}/M_\ast\), where \(M_\mathrm{x}\) 
is the~mass of the~compact companion, and \(\phi=0\) and \(\theta=0\) 
determines the~line-of-centres of the~two stars in the~orbital plane.

   The~radiation pressure in the~continuum caused by the~Thomson
scattering of the~photons emitted by both component stars on free electrons 
is also involved by means of the~Eddington factors \(\Gamma_\ast\) and 
\(\Gamma_\mathrm{x}\), which are chosen as free parameters for
both stars (for which we use the~values 0.26 and 0.31, respectively). 

  These forces cause a~shift in the~critical point of the~outflow that 
varies across the~surface of the~mass-losing star and they distort the~spherical 
symmetry of the~wind. The~mass-loss rate becomes a~strong function of 
the~size of the~primary component relative to its critical potential lobe.
In this radial approximation, we still ignore the~tangential forces
caused by the~pressure gradients between the~neighbouring flux-tubes, as well
as the~Coriolis force, which should be taken into account when computing in 
the~non-inertial reference frame co-rotating with both stars. The~streamlines 
of the~wind are strictly radial and the~material is confined within a~selected
cone, hence the~mass continuity equation becomes 
\begin{equation}\label{MOD6}
\frac{\mathrm{d}\dot{M}}{\mathrm{d}\Omega} = \rho{}\varv{}r^2\; ,
\end{equation}
where \(\mathrm{d}\dot{M}/\mathrm{d}\Omega\) is the~mass-loss rate per unit
solid angle. The~outflow of the~wind is artificially reinforced by
an~unrealistic centrifugal force. We expect this approximation to have
little influence on the~deep highly subsonic layers of the~wind, although
it must lead to a~completely unrealistic model in the~supersonic circumstellar 
region.

%                                                One column figure
%------------------------------------------------ Fig.1 - radwin
\begin{figure}
\centering
   \setlength{\unitlength}{1mm}
 {\includegraphics[width=89.mm]{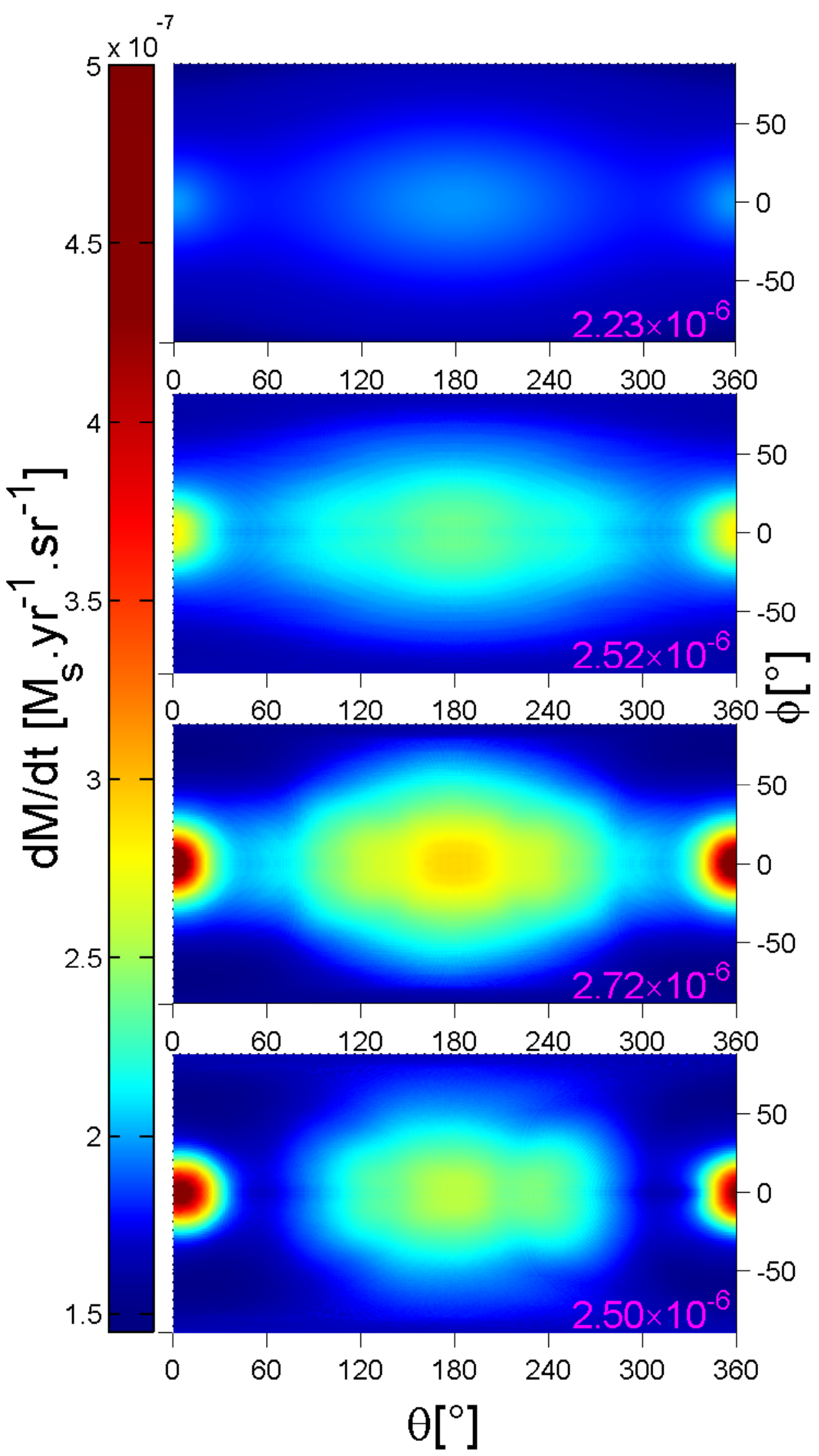}} %{mapy.pdf}}
\caption{ %
Distribution of stellar wind intensity on the~surface of the~optical component
of Cyg X-1 in the~radial approximation (CAK model -- the~uppermost panel),
the~radial approximation of the~radiation-hydrodynamic model in 3D Cartesian
grid (the~second panel), the~non-radial radiation-hydrodynamic solution
without the~Coriolis force (the~third panel) and the~full
radiation-hydrodynamic solution with the~Coriolis
force included (the~bottom panel). The~direction to the~$L_{1}$-point
and the~companion has the~sidereographic longitude 0$^{\circ}$, the~longitude
180$^{\circ}$ is directed towards the~$L_{2}$-point. The~number in the~lower
right corner of each panel gives the~overall outflow of the~material from
the~supergiant in \(M_{\odot}yr^{-1}\)}\label{radwin}
\end{figure}
%---------------------------------------------------------

   The~results of this radial numerical model are presented in 
the~upper panel of Figure~\ref{radwin}, where we display the~directional
dependence of the~stellar wind intensity (i.e. the~mass-loss rate per unit 
solid angle). We have chosen values of these parameters that are appropriate 
for the~high-mass X-ray binary system Cygnus X-1 and its primary O star 
supergiant HDE 226868. The~orbital period is 5.6 days and the~masses of 
the~supergiant and the~black hole have been chosen to be equal to 24. and 
8.7~$M_{\odot}$, resp., within the~still wide region of the~observational 
limits 23$_{-6}^{+8}M_{\odot}$ and 11$_{-3}^{+5}M_{\odot}$ found
by \cite{2009ApJ...701.1895C}. 
These values result in a~distance of 42.4~$R_{\odot}$ between the~component
centres. The~critical Roche lobe of this system has a~mean radius of 
20~$R_{\odot}$: the~$L_{1}$-point is at a~distance of 25.6~$R_{\odot}$, while 
the~intersection of the~critical equipotential with the~rotational axis is 
18.7~$R_{\odot}$ from the~supergiant centre. We have set the~surface of 
the~donor star to a~mean radius 18~$R_{\odot}$ (i.e. at the~equipotential of 
size 19.9~$R_{\odot}$ towards the~companion and 17.3~$R_{\odot}$ towards 
the~pole). The~gravity acceleration $g$ varies across this equipotential from 
1.5 to $3.1\times 10^{3}{\rm cm.s}^{-2}$ towards the~$L_{1}$ and the~pole, 
resp., i.e. the~mean value ${\rm log}\,g=3.3$ is close to the~observational 
limit $3.00\pm0.25$ found by \cite{2009ApJ...701.1895C}. 
Similarly, the~temperature 30.0 kK, which we have taken from the~values 
for which CAK gives the~parameters $k$ and $\alpha$ in Eq.~(\ref{MOD1}),
roughly corresponds to the~value 28.0$\pm2.5\,$kK determined in
\cite{2009ApJ...701.1895C}.

 The~proximity of the~black-hole companion and the~assumed synchronous 
rotation of the~mass-losing supergiant obviously disturb the~spherical
symmetry of the~wind and the~most intensive outflow of the~material is 
concentrated in the~equatorial plane. The~highest intensity can be found 
at a~point directly facing the~black hole. However, more interestingly, 
another significant local maximum is situated on the~completely opposite 
side of the~star. According to our model, the~total mass-loss rate $\dot{M}$ 
of HDE~226868 is close to \(2.23\times10^{-6}M_{\odot}yr^{-1}\). 
The~total outflows from the~facing and opposite hemispheres are almost 
comparable and their ratio depends on the~system parameters. 
These results qualitatively agree with the~earlier works of 
\cite{1982ApJ...261..293F} and \cite{1987PAICz..70..263H}. 
It should, however, be noted that there is a significant quantitative 
difference between the~tidal modulation of the~evaporative and line-driven 
winds. The typical thermal velocity is approximately 22 km/s for 
the~temperature 30\,kK, which, compared to the~critical rotational velocity 
500 km/s, gives a ratio of the~thermal to the~binding potential energy of 
the~order of $10^{-3}$. The~Parker's evaporative wind is thus highly 
concentrated in the~vicinity of the~$L_{1}$-point (cf. 
the~Appendix~\ref{Append2}, Fig.~\ref{OO019f5}). However, if the~CAK- 
radiative drag enhances the~wind and the~radiation  of the~star is assumed 
to be isotropic, the~tidal modulation of the~wind intensity is partly 
smeared out. The distribution of mass-loss rate is then sensitive to 
the~choice of the~surface level.

\section{Radiation hydrodynamic model of the~wind}\label{hydro}
\subsection{Equations of motion}\label{hydroeq}
To deal with the~previously mentioned objection to the~radial approximation, 
we also created a~numerical model of the~stellar wind based on the~radiation 
hydrodynamics calculated in a~3D Eulerian coordinate grid. In this case, 
a new equation of motion is given by 
%\begin{eqnarray}\label{MOD7}	
%	\frac{\partial{}v_x}{\partial{}t}  = &-& v_x\frac{\partial{}v_x}{\partial{}x} - v_y\frac{\partial{}v_x}{\partial{}y} - v_z\frac{\partial{}v_x}{\partial{}z} - \frac{\partial\Phi{}_\mathrm{eff}}{\partial{}x} + (f_L)_x - \frac{1}{\rho}\frac{\partial{}P_\mathrm{g}}{\partial{}x} \nonumber \\
%&+& 2nv_y \nonumber \\
%	\frac{\partial{}v_y}{\partial{}t}  =  &-& v_x\frac{\partial{}v_y}{\partial{}x} - v_y\frac{\partial{}v_y}{\partial{}y} - v_z\frac{\partial{}v_y}{\partial{}z} - \frac{\partial\Phi{}_\mathrm{eff}}{\partial{}y} + (f_L)_y - \frac{1}{\rho}\frac{\partial{}P_\mathrm{g}}{\partial{}y} \; .\\
%&-& 2nv_x \nonumber \\
%	\frac{\partial{}v_z}{\partial{}t}  =  &-& v_x\frac{\partial{}v_z}{\partial{}x} - v_y\frac{\partial{}v_z}{\partial{}y} - v_z\frac{\partial{}v_z}{\partial{}z} - \frac{\partial\Phi{}_\mathrm{eff}}{\partial{}z} + (f_L)_z - \frac{1}{\rho}\frac{\partial{}P_\mathrm{g}}{\partial{}z} 	\nonumber
%		\end{eqnarray}
\begin{equation}
 \frac{\partial\vec{\varv}}{\partial{}t}  = -(\vec{\varv}\vec{\nabla})\vec{\varv}-\vec{\nabla}\Phi_\mathrm{eff}+\vec{f}_{L}-\frac{1}{\rho}\vec{\nabla}P_\mathrm{g}+2\vec{\varv}\wedge\vec{n}\; , %\label{MOD7}\nonumber
\end{equation}
which has to be solved together with the~continuity equation
\begin{equation}\label{MOD8}
 \frac{\partial{}\rho}{\partial{}t} + \nabla\cdot(\rho{}\vec{\varv})=0\; .
\end{equation}
 The constant orbital angular velocity $\vec{n}$ is given
by the~third Kepler's law ($n^2D^3=(1+q)GM_\ast$).
Our present model uses the~purely Newtonian approach to handle 
the~gravity and the~dynamics. We can naturally expect that relativistic 
effects are important in the~close vicinity of the~compact companion 
in the~X-ray binary, which is either a~neutron star or a~black hole. 
However, the~grid we use for our numerical solution is sparse in order 
to achieve a~reasonable computational time. A~typical dimension of one cell 
corresponds to several thousands of Schwarzschield radii and we are unable 
to distinguish structures that are smaller than this typical length. 
It is therefore reasonable to assume that at such a~distance, relativistic 
effects will be negligible.

 We have assumed that the~radiative pressure $\vec{f}_L$ in the~lines, which 
is a~main source of the~repulsive force, is radial and that its value is
also given by Eq.~(\ref{MOD1}). The gravity and continuum radiation 
pressure of both components of the~binary have also been included 
in terms of the~effective potential $\Phi_{\mathrm{eff}}$ given by 
Eq.~(\ref{MOD5}). In addition, new effects are taken into account, such 
as the~tangential components of the~gas-pressure gradient and the~Coriolis 
force. The~gas pressure $P_\mathrm{g}$ was taken from the~equation 
of state of an~ideal gas whose temperature is fixed to the~effective 
temperature of the~star in the~present version of our code. This is 
a~simplifying approximation that we will improve in future
versions of the~model, although it may not be too inaccurate in 
the~case of a~rarefied gas in thermodynamic equilibrium with the~radiation
of both stars. To develop a~self-consistent radiation-hydrodynamical
model, the~radiative transfer should also be solved, whereas in our present
model the~radiation field is prescribed. In the~subsequent paragraph, we
briefly discuss the~possible consequences of this simplification.

\subsection{The radiation field}\label{Radfield}
In his referee report to this paper K. Gayley raised additional objections 
to the~Friend-Castor approximation used to describe the~radial line-driven 
wind in binaries. On the~basis of their investigations of the~stellar 
winds of early-type fast rotators and the~problem of the~(non-)existence of 
wind-compressed disks \citep{1996ApJ...472L.115O,1998Ap&SS.260..149O,
2000ApJ...537..461G} 
Gayley suggested that both the~non-radial components of the~radiative force 
and the~anizotropy of the~supergiant radiation due to the~gravity darkening
may substantialy influence the~structure of the~wind in binaries. To verify
whether this suggestion is indeed valid, we would need to create 
a~sophisticated model that exceeds the~possibilities of our current numerical 
model and intentions of this work. We thus give here only our preliminary 
estimates of the~possible influence of these effects.

 The~non-radial components of the~radiative force in supersonic axisymmetric 
flows were studied by \cite{1978Afz....14..537G}. 
In a differentially rotating medium such as a~Keplerian disk, they are 
due to the~asymmetry of the~velocity gradients in directions skewed forward 
and backward with respect to the~radial direction and the~consequent 
difference in the~depletion or enhancement of the~line radiation 
intensities. 
As for either the~Poynting-Robertson effect \citep{1937MNRAS..97..423R}
or the~reflection of radiation by rotating disks \citep{1997ApJ...480..324H},
this radiative torque can transfer the~energy and angular momentum 
from the~medium to the~radiation and stimulate a~fall down of the~flow. 
\cite{2000ApJ...537..461G} found that this effect may decrease 
the~angular-momentum loss of the~fast rotators by the~wind to about 
30\% or 40\%. 

 However, in the~case of the~synchronously rotating component of a~binary 
with a~non-negligible mass of the~companion, the~rotation is safely 
subcritical (even if the~component fills its Roche lobe, which is restricted 
by the~tides). In the~particular case of our present model, the~rotational 
velocity of the~supergiant is approximately 160 km/s, while the~critical 
velocity is 500 km/s. Consequently, the~influence of the~non-radial 
components of line-driven radiative force is also smaller than 
in the~case of critically rotating stars.

 While these~non-radial components of the~radiative force may to some 
extent influence the~motion of the~circumstellar matter in the~supersonic
region, where a~non-local coupling can also take place, their role is
even more questionable in the~subcritical region where the~surface
distribution of the~wind intensity is determined. In this quite narrow
slab of the~atmosphere, the~tangential component of the~wind velocity
increases only slowly due to the~Coriolis force, hence the~optical depths
in the~skewed directions given by Eq.~(\ref{MOD2}) cannot differ 
significantly.

 A~more substantial objection to the~Friend-Castor
model is the~possible role of gravity darkening. According to
von~Zeipel's theorem \citep[von][]{1924MNRAS..84..665V}, 
the~distribution of the~radiative flux $F$ and the~effective temperature 
$T_{\rm eff}$ across the~tidally or rotationally distorted surface of 
a~star should be given by the~local gravity acceleration $g$
as 
\begin{equation}\label{vonZeipel}
 F\sim T_{\rm eff}^{4}\sim g^{4\beta}\; , 
\end{equation}
where $\beta\equiv 0.25$ in the~standard von~Zeipel's formula. 
Because the~radiative force
$f_{L}$ given by Eq.~(\ref{MOD1}) is proportional to the~flux,
it may enhance the~wind more in the~polar region of the~star,
where $g$ reaches its highest value, than in the~equator and especially
in the~line joining the~component stars, where it has its smallest value.
For the~values of parameters chosen in our calculations, $g$ is
higher by 23\% at the~poles and lower by 15\% and 40\% in 
the~directions towards the~$L_{2}$ and $L_{1}$, resp., than
its value in the~perpendicular direction. This~effect is thus 
the~opposite of the~direct modulation of the~wind by the~effective 
potential. In the~region, where the~distance from the~star is comparable
to its radius, this effect is additionally enhanced by 
the~ellipticity of the~star.

 To determine the~outcome of these mutually competitive effects, one 
would need to develop a~reliable model of the~gravity darkening. 
The~above-mentioned original von~Zeipel's formula is widely used owing 
to its simplicity, but it is neither physically self-consistent nor 
confirmed by observations, which mostly indicate that the~actual gravity 
darkening is smaller, as expected from a~more detailed theoretical 
treatment. \cite{1967ZA.....65...89L} 
reinvestigated the~problem of gravity darkening asuming a~convective 
instead of a~radiative equilibrium arriving to a~generalized formula 
in Eq.~(\ref{vonZeipel}) in which the~exponent $\beta=0.08$. 
The~classical von~Zeipel's formula is based on the~solution of 
radiative transfer in the~diffusion approximation only, assuming 
hydrostatic equilibrium and hence the~homogeneity of the~atmosphere 
on equipotential surfaces (cf. Appendix~\ref{Append1}). 
The~simplified treatment of radiative transfer neglects both the~dilution 
of the~radiative flux in the~parts of sub-surface layers with higher 
curvature, which enhances the~gravity darkening, and the~tangential 
diffusion of the~radiation, which decreases its effect,  
\citep[cf.][]{1992A&A...256..519H}. 
However, even more importantly, the~resulting gravity darkening violates 
the~homogeneity and the~hydrostatic equilibrium, which are the~starting 
assumptions of von~Zeipel's theorem, and gives rise to meridional 
circulations. The Coriolis force acting on these flows may then also break
the~mirror symmetry of the~advancing and recessing hemispheres of the~star 
and hence the~simple picture of the~flux given by the~local value of 
\(g\) \citep[cf.][]{2007ASPC..367..393G}. 
Our detailed solution of radiative transfer in plane-parallel 
grey atmosphere in exact hydrostatic equilibrium yields $\beta\simeq 0.13$ 
for small variations in $g$ and a~more complex dependence $F=F(g)$ for
its larger variations \citep[cf.][and Appendix~\ref{Append1}, 
particularly Figure~\ref{OO019f2}]{1987PAICz..70..263H}.
Yet another value $\beta\simeq0.19$ was obtained by
\cite{2012A&A...538A...3C}, who also gives references to several
observational results confirming that the~actual gravity darkening
parameter $\beta$ has a~scatter and is usually smaller
than the~von~Zeipel's value of 0.25.

%------------------------------------------------ Fig.2 - gravdark
\begin{figure}
\centering
   \setlength{\unitlength}{1mm}
 {\includegraphics[width=89.mm]{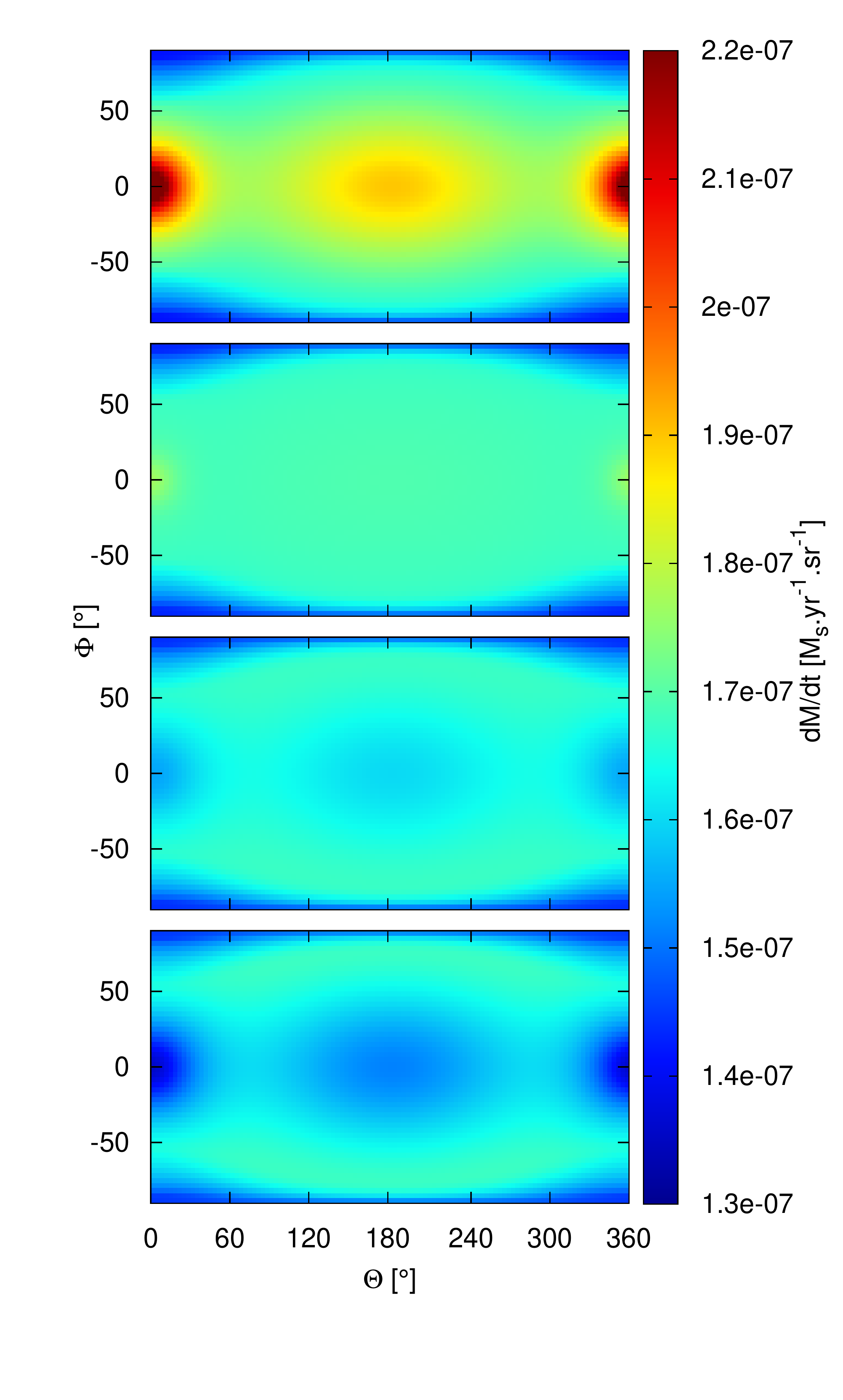}} %{mapy.pdf}}
\caption{ %
Distribution of the~stellar wind intensity in dependence on gravity 
darkening. The~radial approximation for the~same parameters as in 
Fig.~\ref{radwin} is calculated for radiation modulated according to 
Eq.~(\ref{vonZeipel}) and values of the~parameter $\beta$ from the~top 
to the~bottom of 0.0, 0.125, 0.1875, and 0.25 
}\label{gravdark}
\end{figure}
%---------------------------------------------------------

 On the~referee's suggestion, we recalculated the~outflow in the~radial
approximation described in Section~\ref{radwind}, assuming the~gravity
darkening given by Eq.~(\ref{vonZeipel}) with several values of 
the~parameter $\beta$ (cf. Fig.~\ref{gravdark}). 
For the~medium value $\beta=0.125$, the~tidally enhanced outflows in 
the~directions of the~$L_{1}$- and $L_{2}$-points can really be diminished
due to the~decrease of the~radiative drag. For even higher values of $\beta$, 
our calculations show a~formation of local minima of the~wind intensity in 
these directions, although the~local minima caused by the~potential barrier 
in polar regions persist and the~maximum of the~wind intensity appears at
moderate latitudes. This is, however, an upper estimate based on a~model 
with several inconsistencies.

 Because the~starting assumption of hydrostatic equilibrium is violated 
from the~very beginning in the~case of strong stellar winds, we have not 
included the~von~Zeipel's formula for gravity darkening in
our 3D models. To achieve a really self-consistent radiative-hydrodynamic
solution, the~3D radiative transfer should be included in the~procedure,
which would be computationally much more demanding even in the~simpler case
of axially symmetric rotating stars. In the~current approximation,
we include in the~hydrodynamic modelling the~radiative forces from
the~given purely radial and isotropic radiation of the~components 
and postpone the~treatment of
radiative transfer in the~circumstellar matter to a~future study.

\subsection{Numerical model}
   The~results of the~previously computed radial model, i.e.
the~spatial distribution of density $\rho$ and the~Cartesian components 
of the~velocity vector $\vec{\varv}$, were used as initial conditions for  
the~corresponding quantities in the~new 3D hydrodynamic simulation 
of the~time evolution of the~wind and circumstellar matter. They were also
used to define the~inner boundary conditions for the~gas outflow in 
the~photosphere of the~donor star, which is nearly in hydrostatic equilibrium
and the~effects of the~additional terms could thus be neglected there.

   An~equidistant Cartesian coordinate grid was chosen in the~co-rotating
reference frame. The~values of $\rho$ and $\vec{\varv}$ found from 
the~radial-wind model were interpolated onto this grid and kept fixed inside 
a~Roche equipotential chosen to correspond to the~surface of the~donor star 
in order to define the~inflow boundary condition for the~Eulerian 
radiation-hydrodynamic solution. The~outer boundary condition at the~edge 
of the~grid is a~simple outflow boundary condition with derivatives of all 
variables set equal to \(0\). The time derivatives of the~density and 
components of the~velocity vector were calculated from the~equations of 
the~mass and momentum conservation, resp. The~small corrections 
corresponding to a~sufficiently short time-step \(\Delta{}t\) were then 
added to the~initial values. Repeating this process, the~initial stationary 
solution of the~radial model adapted to satisfy the~conditions of the~new 
model. A~new stationary solution was achieved after a~relatively quick 
convergence of all quantities. The~sum of time steps typically corresponds 
to a~small number of orbital periods of the~system (cf. Figure~\ref{evol01}). 
This is obviously a~consequence of the~wind velocity at the~critical point 
given by the~sound speed $a$ being comparable to the~orbital velocity $nD$ 
in our investigated cases when the~mass loss by stellar winds is substantial. 
The~convergence could be much slower in the~sub-sonic layers of the~wind, 
if the~inner boundary conditions were chosen on a~deep equipotential
where the~density is high and outflow velocity low. However, to avoid 
the~steep density gradients in these layers, which could not be represented
with sufficient accuracy in the~limited space-resolution of the~grid, we
had to choose this layer closely below the~surface of critical points.
The~convergence of the~numerical model was thus accelerated by the~numerical
viscosity within a~few grid cells around the~inner boundary surface.

   A~value of time step \(\Delta{}t\), which is constant for the~whole grid,
was subjected to the~Courant--Friedrichs--Lewy condition, specifically
\begin{equation}\label{MOD9}
\Delta{}t =C_0\mathrm{min}\left(\frac{1}{a}\mathrm{min}\left(\Delta{}x,\Delta{y}, \Delta
{}z\right),\frac{\Delta{}x}{|\varv_x|},\frac{\Delta{}y}{|\varv_y|},\frac{\Delta{}z
}{|\varv_z|}\right)\; ,
%\Delta{}t = C_0\mathrm{min}\left(\frac{1}{a}\mathrm{min}\left(\Delta{}x,\Delta{}y\right),\frac{\Delta{}x}{|v_x|},\frac{\Delta{}y}{|v_y|}\right)\; ,
\end{equation} 
where \(\Delta{}x\), \(\Delta{}y\), and \(\Delta{}z\) are the~distances 
between the~neighbouring grid nodes in the~\(x\), \(y\), and \(z\)-directions, 
resp. The~Courant number \(C_0\) is a safety factor, which in this case 
is taken to be \(0.5\). 

   We chose a~combination of the~forward-time central-space scheme (FTCS 
hereafter) and the~Lax-Friedrichs numerical integration scheme in our 
computations.  This combination allowed us to exploit all of the~advantages
of both schemes, in addition to suppressing most of their drawbacks. 
The~FTCS is a~very simple and computationally fast integration scheme, 
which is unfortunately always unstable.  To overcome this instability, we 
introduced a~correctional mechanism to prevent random numerical peaks. 
On the~other hand, the~Lax-Friedrichs scheme is stable in all cases when 
the~Courant stability condition is satisfied and therefore it serves as 
a~stabilizing component in our model.  However, the~Lax-Friedrichs scheme 
is characterized by a~high numerical viscosity, which results in a~high 
level of diffusion. For a~large number of integration steps, this feature 
would lead to smearing of all steep gradients, even those that are physically 
interesting.  This is why we applied the~Lax-Friedrichs scheme as infrequently
as possible, specifically only once in every 25 integration steps.  If we 
had included the~Lax-Friedrichs scheme with a~lower frequency, the~numerical 
simulation would have become unstable in the~long term. 

   The~additional averaging of the~density and the~velocity in the~outer 
boundary of the~integration area was added to prevent oscillations and 
numerical instabilities within the~stationary solution. This averaging is
responsible for less accurate results in this area. Nevertheless, the
density is very low at the~border of the~integration area and therefore
should not have a~measurable impact on the~solution in the~inner regions. 

  The~evolution of the~density distribution in the~equatorial plane 
during the~convergence process towards a~new stationary solution is 
captured in Figure~\ref{evol01}. The~initial state corresponds to 
the~conditions set up by the~radial approximation. The~integration process 
was stopped when a~nearly steady-state solution corresponding to the~newly 
included forces -- the~Coriolis force and the~tangential forces caused 
by the~pressure gradients -- was reached. The~white index in the~upper 
right corner of each panel indicates an~evolutionary time expressed in 
the~units of the~orbital periods. The~red cross and the~asterisk show 
the~position of the~black hole and the~centre of the~supergiant, resp. 
The~patch at the~centre of each panel represents the~supergiant component 
of the~binary. More specifically, it is an intersection of the~volume 
defined by an~equipotential surface and the~equatorial plane. 
The~convergence to a~new stationary solution is relatively quick: at 
the~time index equal to 3.0, the~changes of studied quantities in one
time step are less than \(10^{-4}\) of their initial values. As indicated 
in Figure~\ref{evol01}, the~originally radial distribution of density 
becomes strongly influenced by the~presence of the~compact companion. 
We can see a~gradual formation of a~density tail behind the~compact 
component, which is diverted by the~Coriolis force. It is also apparent 
that the~density increases with time throughout the~entire modeled space. 
The~reason for this increase is that we reduced 
the~centrifugal force by allowing the~wind to be diverted by the~Coriolis 
force. Therefore, the~repulsive net force is weakened, leading to 
the~slowdown of the~radial component of the~wind velocity and hence due to
the~equation of mass continuity also to the~increase in the~wind density. 

%------------------------------------------------ Fig.3 - evol01
\begin{figure}
\centering
   \setlength{\unitlength}{1mm}
 {\includegraphics[width=89.mm]{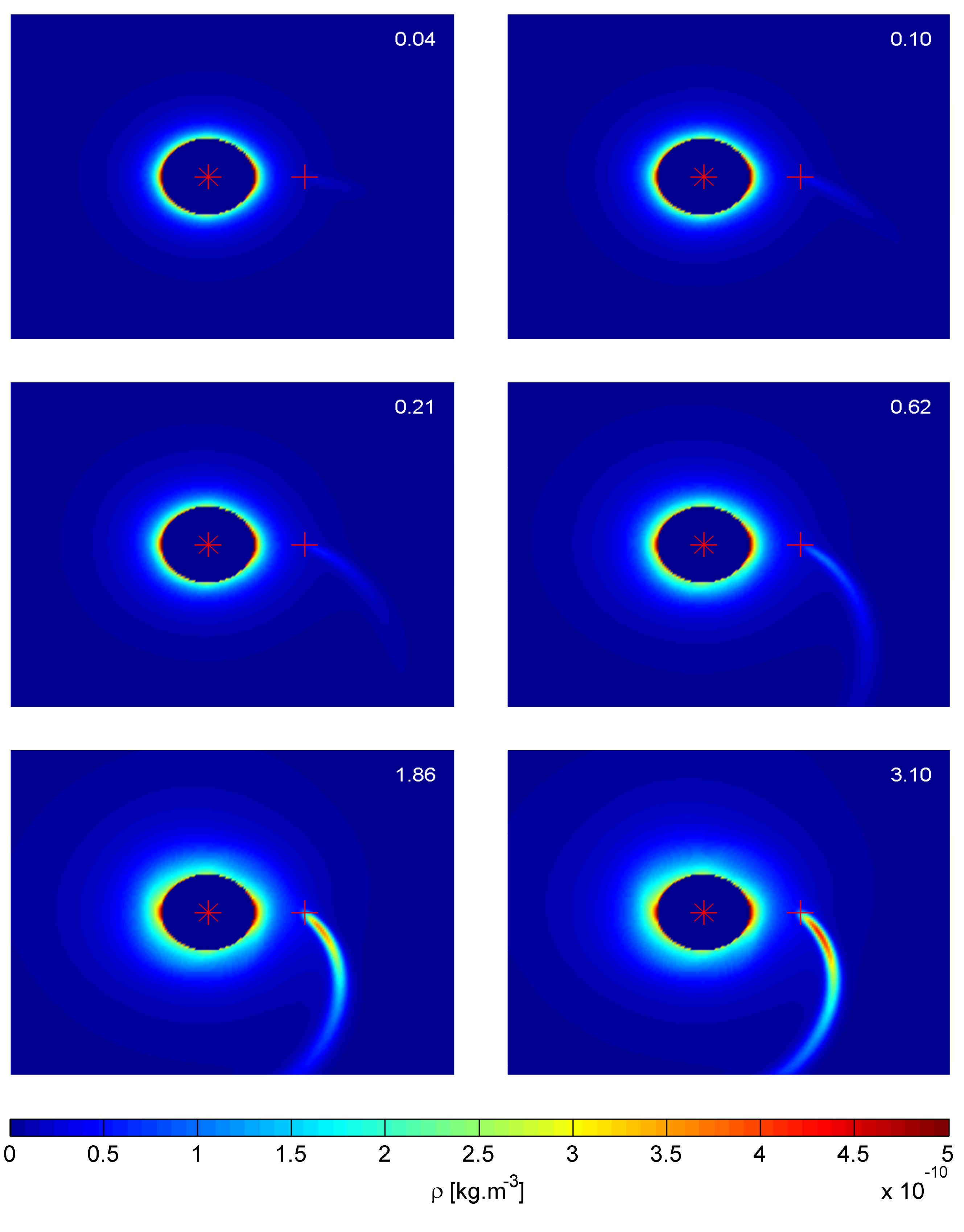}}
\caption{ %
Evolution of the~density distribution in the~equatorial plane computed with 
the~radiation hydrodynamic model during the~convergence process towards a~new
stationary solution. The~uppermost left panel represents the~state close to 
the~conditions set up by the~radial model. The~bottom right panel captures 
the~nearly steady-state solution corresponding to the~newly included forces 
-- the~Coriolis force and the~tangential forces caused by the~pressure 
gradients. The~white index in the~upper right corner of each panel indicates 
an~evolutionary time expressed in the~units of the~orbital periods. 
The~red cross and asterisk show the~position of the~black hole and the~centre 
of the~supergiant, resp.}\label{evol01}
\end{figure}
%---------------------------------------------------------

   A~modified directional distribution of the~stellar-wind intensity is
traced back to the~roots of the~streamlines on the~Roche equipotential 
surface at the~centre, which was selected to represent a~surface of 
the~supergiant. The~result for the~same stellar parameters corresponding 
to the~system Cyg X-1 is shown in the~bottom panel of Figure~\ref{radwin}. 

   To distinguish the~role of the~different numerics and the~physical
effects involved in the~new model, we also calculated two artificial
intermediate models: the~radial model in the~3D Cartesian grid and
a~non-radial model with tangential gradients and velocity components
included but for which the~Coriolis force had been neglected (cf. the~second and the~third 
panel of Figure~\ref{radwin}, resp.).  

   The~second panel of Figure~\ref{radwin} represents results of the~3D
hydrodynamical simulation where the~non-radial components of velocity were
suppressed in each step of the~computing 
process. We thus allowed the~wind material to move in the~radial direction 
only and forced it to reach a stationary solution independently in each 
radial cone. Consequently, this solution should correspond to the~radial
approximation depicted in the~uppermost panel.  However, the~numerical 
results show a~considerably higher intensity of the~stellar 
wind all across the~surface of the~supergiant with a particularly enhanced
outflow in the~direction of the~$L_{1}$-point and the~integrated outflow
increase to \(2.52\times10^{-6}M_{\odot}yr^{-1}\). 
This difference between the~two physically equivalent models is obviously
due to numerical errors caused by the~insufficient spatial resolution of 
the~3D-grid in the~inner region of the~computing area where we meet a~steep 
density gradient. As a~result, we experience an~increase in the~density all 
across this region and consequently a~higher level of intensity of the~stellar 
wind. To achieve certain qualitative conclusions about the~role of 
the~non-radial motion of the~wind and the~Coriolis force in real binaries, we 
have to compare the~subsequent results of our 3D hydrodynamical model with
this normalized radial approximation solution. 

   In the~third panel of Figure~\ref{radwin}, the~results of the~midway 
hydrodynamic simulation are presented. In this model, we also take into 
account the~tangential forces caused by the~pressure gradients between 
the~neighbouring flux-tubes. We set no additional conditions on the~velocity 
so the~streamlines of the~wind can bend into non-radial directions though 
the~Coriolis force is still excluded. We note that the~overall intensity of 
the~wind has increased compared to the~normalized radial approximation 
solution in the~second panel. The~directional distribution of the~outflow 
has also changed. The~intensity in the~equatorial plane has increased, in 
contrast to the~drop-off that has occured in the~polar regions. Allowing 
the~non-equatorial streamlines to deviate from the~radial direction, they 
can perceive the~centrifugal force and are slightly bent towards the~plane 
of the~orbital motion of the~binary. The~most significant increase in 
the~outflow at the~primary maximum happens because the~streaming of gas 
through the~Roche potential window around the~$L_{1}$-point is more effective 
if non-radial velocities are allowed. A~larger amount of material is 
transported in the~direction with the~lower effective potential barrier and 
the~overall intensity of the~outflow increases to 
\(2.72\times10^{-6}M_{\odot}yr^{-1}\).

\subsection{Results of the computations}
   The~final solution of the~3D hydrodynamical model including 
the~Coriolis force is shown in the~bottom panel of Figure~\ref{radwin}.
Comparing it with the~results of the~previously described model in the~third 
panel we note a~slight shift in the~primary and secondary maxima 
counter-clockwise in the~equatorial plane. This effect can naturally be
expected if we introduce the~Coriolis force into our problem -- most of
the~matter escapes along the~streamlines that pass through the~$L_{1}$ or 
$L_{2}$ points or their vicinities and originate in deeper layers shifted 
in the~direction of the~orbital rotation. Another important feature of 
the~new stationary solution is that the~mass-loss rate at the~primary 
maximum is lower than the~value of the~same quantity in the~previous model. 
We note that there is an~even more significant decrease in the~intensity at 
the~secondary maximum on the~opposite hemisphere from the~black-hole 
companion. The~overall intensity has decreased and the~integrated mass-loss 
rate for the~entire star has dropped to \(2.50\times10^{-6}M_{\odot}yr^{-1}\). 
This effect is also caused predominantly by the~Coriolis force -- the~flux
tubes originating in parts of the~stellar surface where the~gravity is most 
significantly reduced by 
the~tidal force must overcome a higher barrier of the~Roche potential,
while the~actual maximum of the~outflow around the~Lagrange points is less 
enhanced by the~tides. In other words -- we now allow streamlines of  
the~stellar wind to curve and not to co-rotate with the~orbital motion. 
In consequence, we reduce the~centrifugal force, which repulses the~material
from the~star, and therefore we may expect that the~overall mass-loss rate
be lower than in the~hydrodynamical simulation in which the~Coriolis force 
is not included, as predicted by S. K\v{r}\'{\i}\v{z}. The~effects of the~non-radial
flowing and the~Coriolis force practically cancel each other in our present
calculation. However, they may be in a different proportion in other systems
and with the~temperature structure taken into account. The~classical models
of radial stellar winds in binaries should thus be revised to achieve reliable
quantitative results.

%                                                One column figure
%--------------------------------------------------------Fig.4 proud
\begin{figure}
 \centering
 \includegraphics[width=8.6cm]{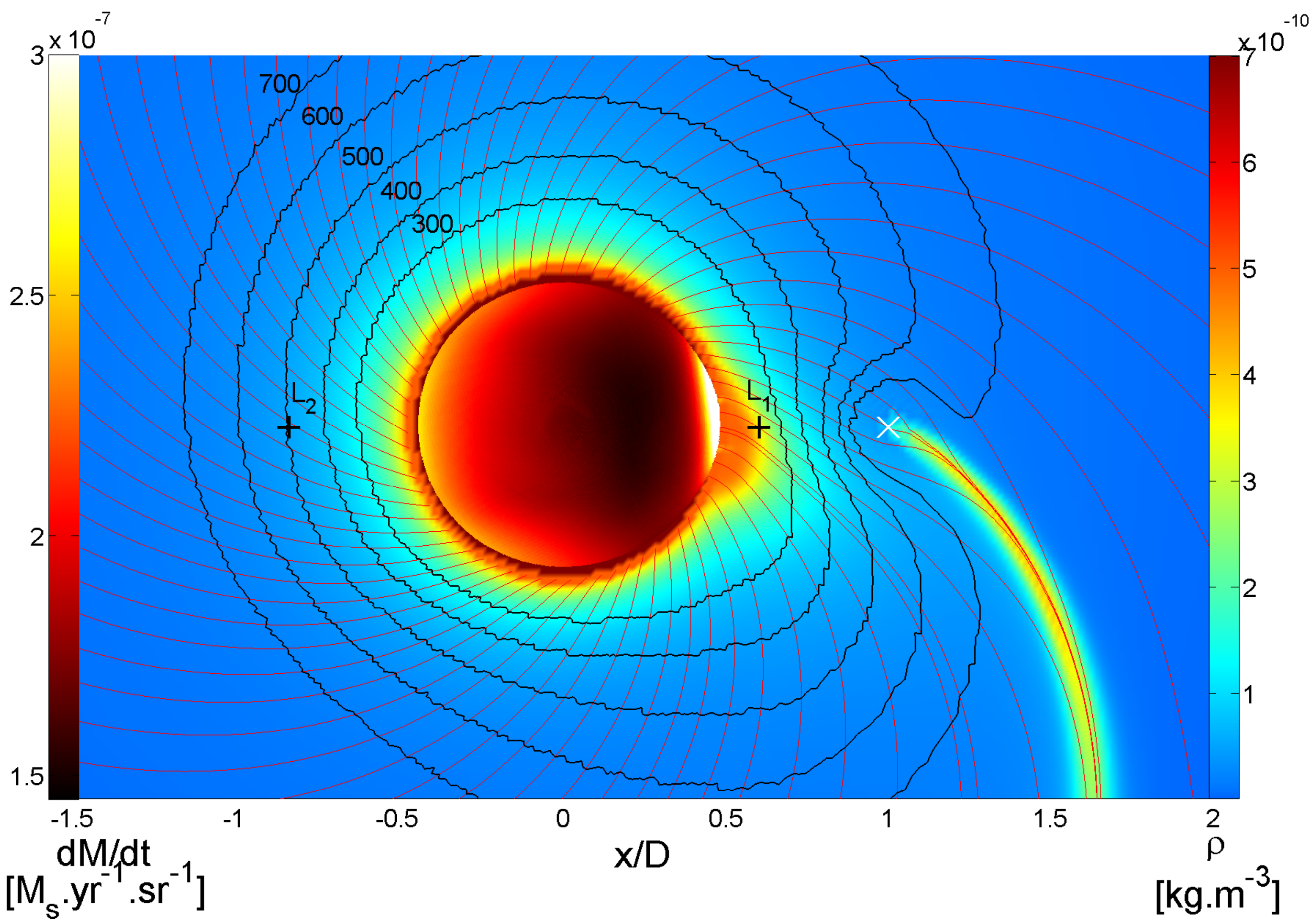} %{mix.pdf}
   \setlength{\unitlength}{1mm}
\caption{ Hydrodynamic solution of the~wind-fed circumstellar matter. 
The~white cross shows the~position of the~black-hole companion, the~black 
crosses mark the~positions of the~points $L_{1}$ and $L_{2}$.  The~distance 
scale on the~x-axis is labelled in units of the~separation between both 
components of the~binary.  The~right-hand-side colour-scale corresponds 
to the~distribution of the~wind density in the~equatorial plane. 
The~left-hand-side colour-scale describes the~directional dependence of 
the~intensity of the~stellar wind projected onto the~surface of 
the~supergiant.  The~red lines represent the~streamlines of the~stellar 
wind and the~black contour-lines labelled by the~corresponding 
numbers in \(km\cdot{}s^{-1}\) mark the~absolute values of 
the~stream velocity in the~orbital plane}
 \label{proud}
\end{figure}

   The~flow of the~circumstellar matter in the~wider vicinity of the~
mass-losing star can be seen in Figure~\ref{proud}, which represents 
the~density, velocity, and streamlines in the~orbital plane. The~colour
map on the~projection of the~tidally distorted disk of the~star reflects 
the~wind density shown in Figure~\ref{radwind} projected from the~direction 
of the~rotational axis (its minimum in the~polar region and the~maxima shifted 
with respect to the~region below the~Lagrangian points can be clearly seen). 
The~colour-map outside the~stellar surface represents the~density distribution,
the~approximately concentric isocontours show the~values of the~velocity,
while the~radial curves show the~streamlines in the~equatorial plane.  
In general, the~density 
of the~stellar wind decreases while the~velocity increases 
with the~distance from the~supergiant. The~distortion of the~streamlines
by the~Coriolis force is obvious and its role in bringing the~mass
ejected around the~$L_{2}$-point into the~vicinity of the~compact companion
can be estimated. (The~significance of this effect argued by D. Gies
depends on the~parameters of the~system and the~regime of the~radiative 
acceleration).

   The~most spectacular feature in this Figure~\ref{proud}, is the~dense 
tail that appears behind the~compact companion (the~position of 
the~companion is marked by the~cross). The~growth of this tail can be 
seen in Figure~\ref{evol01}. Its source is 
obvious from the~streamlines of the~wind. The~flow of the~wind is 
focused by the~intense gravity field of the~black hole, and the~mass density 
increases in the~tail as the~streamlines passing the~black hole at larger 
distances join it. This effect, which is naturally beyond of scope of 
the~models of radial wind, is obviously a~version of the~Bondi-Hoyle-Lyttleton 
(BHL hereafter) accretion (with differences described in the~next Section). 
In an~approximation of free-particle motion, this effect is analogous 
to the~gravitational focusing of photons and the~formation of high-density 
caustics. Our isothermal model facilitates the~condensation of the~tail, 
although a~similar increase in the~mass density behind the~companion star can 
be seen in the~results of many numerical models of mass exchange in binaries.

   An~important characteristic of the~BHL accretion investigated already 
in the~original papers of its discoverers but often neglected in more recent 
papers is its instability.  There is no unique steady-state solution with 
a~strict boundary between the~region of the~tail that falls on the~accretor 
and which escapes to infinity. The~accumulation of the~mass in 
the~tail and the~variability of the~accretion rate with time may play 
an~important role in triggering the~transitions of X-ray binaries between 
the~radiatively high/soft and low/hard regimes of the~accretion. Our 
computations also indicate that there are some instabilities in the~tail,
although an~increase in the~spatial resolution and a~sophistication of 
the~physical model will be needed to verify or disprove these results.

   The~close proximity of a~luminous X-ray source could influence 
the~dynamics of the~line-driven wind by producing a~change in 
the~temperature and 
ionization structure of the~wind material. The~X-ray heating and 
photoionization can inhibit the~wind acceleration by reducing population
of electron levels available to absorb the~momentum of the~radiation
flux from the~primary. In the~case of a~very strong X-ray flux, 
the~photoionization radius can reach the~photosphere of the~primary 
causing the~line-driven wind to never achieve the~escape velocity. 
It may then cease to supply the~accretion process with material. 
Although the~calculations of the~ionization structure were not included 
in the~present version of the~code, we were able to simulate 
the~above-mentioned effect by the~assumption that the~\(k\) parameter
from Eq.~(\ref{MOD1}) is artificially set to 0 for the~wind material, which
leaves the~part of the~photosphere facing the~X-ray source. If 
the~CAK line-driven mechanism is a~main source of the~repulsive force 
acting on the~wind material, by cancellation of the~line force, the~wind 
material loses its support and starts falling back onto the~primary. 
The~situation is captured in Figure~\ref{evol02}. The~initial state 
(the~uppermost left panel) corresponds to the~steady-state solution 
depicted in Figure~\ref{proud}. There is an~apparent steep increase in
the~density in the~region where the~line force has been artificially switched 
off. This is caused by the~slowdown of the~wind velocity in this region 
and also by the~higher layers of the~wind material failing to achieve 
the~escape velocity and falling back onto the~photosphere.

%------------------------------------------------ Fig.5 - evol02
\begin{figure}
\centering
   \setlength{\unitlength}{1mm}
 {\includegraphics[width=89.mm]{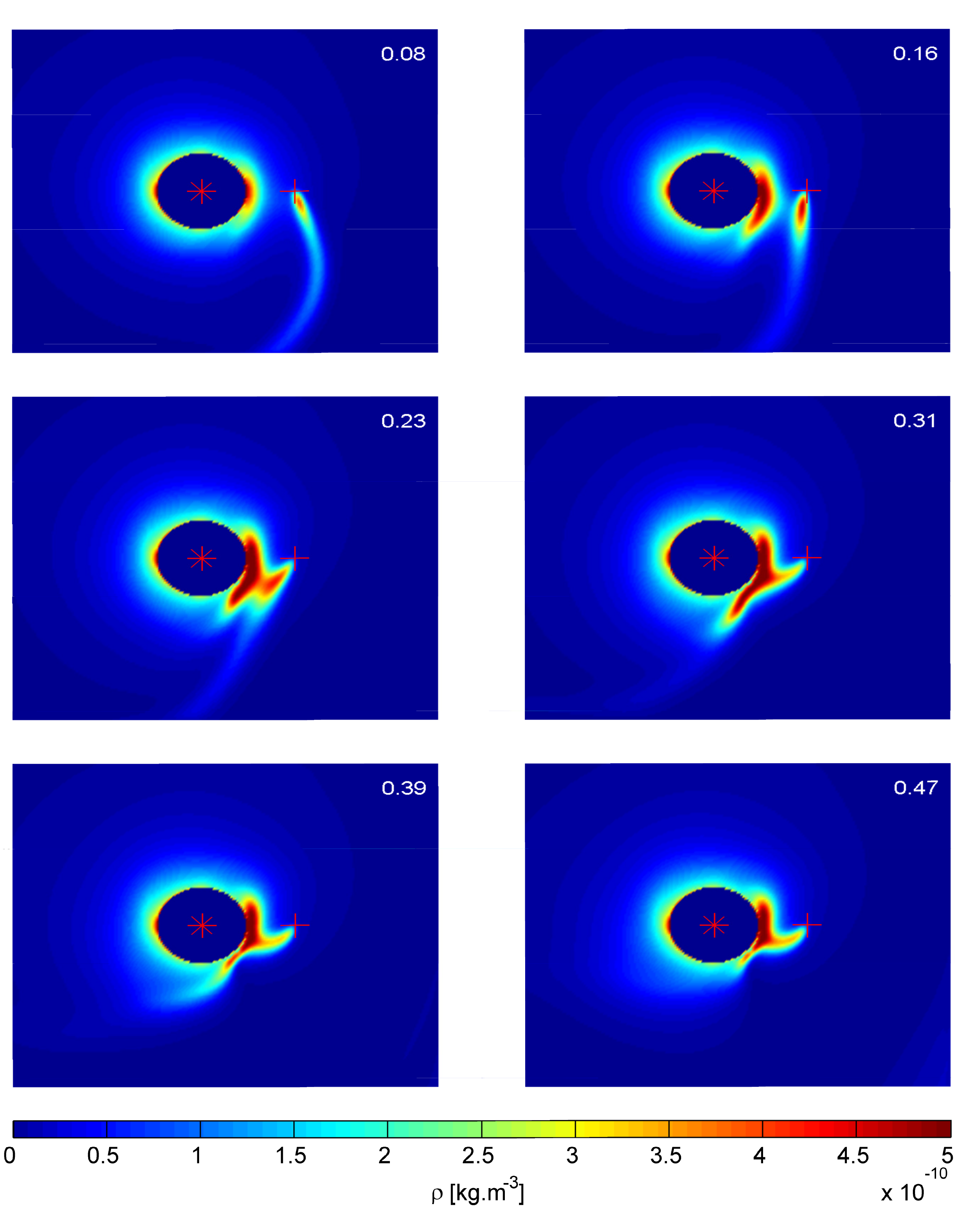}}
\caption{ %
Evolution of the~ density distribution in the~equatorial plane computed 
with the~radiation hydrodynamic model after the~line-driven force from 
the~companion-facing hemisphere was artificially switched off. The~initial 
state (the~uppermost left panel) corresponds to the~steady-state solution 
depicted in Figure~\ref{proud}. The~bottom right panel captures the~new 
nearly steady-state solution. The~white index in the~upper right corner 
of the~each panel indicates an~evolutionary time expressed in the~units 
of the~orbital periods. The~red cross and asterisk show the~position of 
the~black hole and the~centre of the~supergiant, resp.}\label{evol02}
\end{figure}
%---------------------------------------------------------

\section{Bondi-Hoyle-Lyttleton accretion}\label{BHL}
The~standard theory of BHL accretion assumes that the~accreting object 
moves through a~homogeneous static gas (cf. \cite{1939PCPS...34..405H}, 
\cite{1944MNRAS.104..273B}, or \cite{2004NewAR..48..843E}
for a~review). This means that the~initial velocities of the~incident
gas elements are parallel, while the~stellar wind diverges. This difference
is essential because the~divergence limits the~mass rate of the~accretion
into the~formed tail \citep[cf. ][]{1984BAICz..35..343H}. 
Moreover, for an~accretor moving in either a~perpendicular or skewed 
direction with respect to the~source of the~wind, the~axial symmetry of 
the~flow is broken.

   To modify the~calculation of BHL accretion in the~case of 
the~gravitational interaction of a~radial stellar wind with an~orbiting 
satellite, the~boundary conditions must be modified. We define
\begin{equation}\label{BH1}
 w^{i}(r)=w(r)\frac{r^{i}}{r}
\end{equation}
to be the~unperturbed velocity of the~wind and $R^{i}(t)$ to be the~track 
of the~satellite.  The~relative velocity of the~wind in the~instantaneous
inertial system of the~satellite can be expanded in variable $x\equiv r-R$ 
up to linear terms as
\begin{equation}\label{BH2}
 \varv^{i}(x)\equiv w^{i}(R+x)-\dot{R}^{i}=\varv_{0}^{i}+a^{i}_{j}x^{j}\; ,
\end{equation}
where
\begin{equation}\label{BH3}
 \varv_{0}^{i}=w(R)\frac{R^{i}}{R}-\dot{R}^{i}
\end{equation}
is the~velocity of the~wind in the~rest frame of the~satellite colliding
with it head-on, and
\begin{equation}\label{BH4}
 a^{i}_{j}= \frac{w(R)}{R}\delta^{i}_{j}
  +\frac{\mathrm{d}}{\mathrm{d}R}\left(\frac{w(R)}{R}\right)\frac{R^{i}R^{j}}{R}
\end{equation}
is a~symmetric matrix describing the~dilution of the~unperturbed wind
in the~space around the~satellite with respect to its rest frame.
Its projection to the~plane perpendicular onto $\varv_{0}$
\begin{equation}\label{BH5}
 A^{k}_{l}\equiv P^{k}_{i}a^{i}_{j} P^{j}_{l} = \frac{w(R)}{R}P^{k}_{l}
  +\frac{\mathrm{d}}{\mathrm{d}R}\left(\frac{w(R)}{R}\right)P^{k}_{i}\frac{R^{i}R^{j}}{R} P^{j}_{l}\; ,
\end{equation}
where the~projector
\begin{equation}\label{BH6}
 P^{i}_{j}= \delta^{i}_{j}-
  \frac{\varv^{i}_{0}\varv^{j}_{0}}{\varv^{2}_{0}}
\end{equation}
yields the~apparent divergence of the~wind streamlines. For a~vector $z$
that is perpendicular to the~orbital plane (i.e. $(zR)=(z\dot{R})=
(z\varv_{0})=0$ and consequently $Pz=z$), we get
\begin{equation}\label{BH7}
 A^{k}_{l}z^{l} = \frac{w(R)z^{k}}{R}\; ,
\end{equation}
which means that the~dilution of streamlines in this plane corresponds to 
the~proper distance $R$ of their source. On the~other hand, for a~vector 
$y$ in the~orbital plane perpendicular to $\varv_{0}$, i.e.
\begin{equation}\label{BH8}
 y^{l} =c [R^{l}(\dot{R}^{2}-\frac{w(R)}{R}(R\dot{R}))+\dot{R}^{l}(Rw(R)-(R\dot{R}))],
\end{equation}
(for which $(y\varv_{0})=0$ and hence $Py=y$, but $(y\dot{R})=(yR)\frac{w(R)}{R}
=\frac{cw(R)}{R}[R^{2}\dot{R}^{2}-(R\dot{R})^{2}]\neq 0$),
we get
\begin{eqnarray}\label{BH9}
 A^{k}_{l}y^{l}& =&\\
&&\hspace*{-5mm} =\left[\frac{w(R)}{R}
 +\frac{R^{2}\dot{R}^2-(R\dot{R})^{2}}{Rw^{2}(R)-2(R\dot{R})w(R)+R\dot{R}^{2}}
 \frac{\mathrm{d}}{\mathrm{d}R}\left(\frac{w(R)}{R}\right)\right]y^{k}\; .\nonumber
\end{eqnarray}
In particular, when the~satellite moves perpendicularly
to the~wind (i.e. $(R\dot{R})=0$) and the~wind is not accelerated
($\frac{\mathrm{d}}{\mathrm{d}R}\left(w(R)\right)=0$), the~apparent divergence of streamlines in the
orbital plane is reduced as if the~distance of their source is
increased by the~square of the~ratio of the~apparent ($\varv_{0}$) to
the~true ($w$) wind velocity
\begin{equation}\label{BH10}
 A^{k}_{l}y^{l} = \frac{w^{3}(R)y^{k}}{\varv^{2}_{0}R}\; .
\end{equation}

%                                                One column figure
%--------------------------------------------------------Fig.6 obr1
\begin{figure}
 \centering
   \setlength{\unitlength}{1mm}
\includegraphics[viewport=90 115 295 220,width=8.6cm]{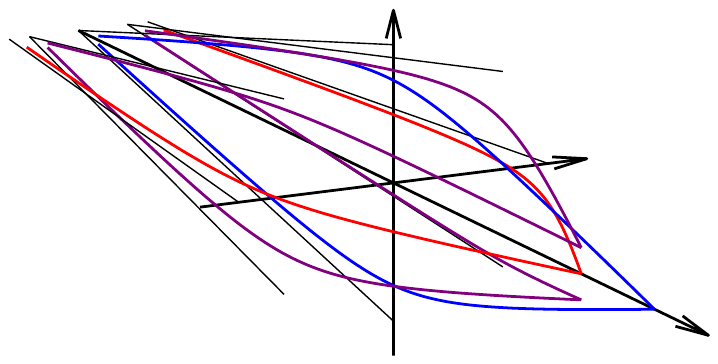}%{BHLaa.pdf}
 \put(-5,7){$x$}
 \put(-20,27){$y$}
 \put(-37,40){$z$}
\caption{\label{BHLfig} Gravitational focusing of the~wind in the~satellite's
co-moving frame in the~approximation of free-particle motion.  The~thin
black lines show asymptotes of the~Keplerian hyperbolic trajectories.
Owing to an aberration of the~initial velocities, the~trajectories 
with the~same impact-parameter in the~orbital plane $(x,y)$ (red in the
electronic version) reach the~$x$-axis closer to the~gravitating
body (placed in the~coordinate centre) than those in $(x,z)$-plane
(blue lines).  The~particles moving out of these planes (violet lines)
collide mutually in the~$(x,z)$-plane before reaching the~$x$-axis} 
 \label{obr1}
\end{figure}

   As a~consequence, the~streamlines in the~equatorial plane ($x,y$ in
Figure~\ref{BHLfig}) are focused on the~$x$-axis more efficiently than
the~streamlines in the~perpendicular plane ($x,z$).  The~streamlines
passing the~satellite out of both these planes cross the~$x,z$-plane
earlier than the~orbital plane and are thus not focused exactly on
the~$x$-axis; the~focusing is astigmatic.  This conclusion agrees with
the~results of numerical models described in the~previous Section, which
show that the~cross-section of the~tail is elongated in the~direction
perpendicular to the~orbital plane.  The~more complicated structure
of the~tail will thus need to be treated using detailed 3D hydrodynamic 
modelling to study the~time variability of the~accretion rate.

\section{Discussion and conclusions}\label{dis}
The~increasingly high quality data available owing to improved observational
techniques at various wavelengths now enable the~more detailed investigation 
of the~circumstellar matter in the~interacting binaries.  An~accurate
interpretation of these observations 
requires more sophisticated theoretical models to get more precise
predictions of the~observable quantities and test observationally the
viability of the~models and their underlying assumptions.

   To yield physically self-consistent and quantitatively reliable
results, our numerical model of the~3D radiation hydrodynamics of stellar 
winds in high-mass X-ray binaries 
\citep[cf.][for a recent review]{2010ASPC..422...57N}
presented here needs to be improved in terms of several aspects of 
physics (first of all to model the~radiative transfer of the~X-rays from 
the~companion and their interaction with the~wind) and numerical
properties (higher resolution).  However, even in its present form 
it shows that non-radial models are necessary to derive quantitative 
information about the~angular modulation of the~wind and the~mass-loss 
and accretion rates with an~accuracy of about ten percents.  
The~preliminary results of our calculations in which the~radiative pressure 
in lines is scaled to account qualitatively for the~ionization of the~wind 
by X-ray illumination in different states of the~compact companion show 
the~significant non-linear response of the~wind strength to
the~radiative drag. 

   The~formation of narrow gaseous tails (or `wakes') behind compact objects 
can be seen in our numerical hydrodynamic models of high-mass binaries. 
Similar results have also been found in several previous hydrodynamic
simulations of isothermal flows -- e.g. \cite{1993ApJ...404..706I},
while the~adiabatic flows tend to form wider shock cones -- cf. 
\cite{1991A&A...248..301M}.  (Note that the~equations of motion (3) and 
(2.2), resp., are incorrectly typed in these papers.)
The~physics of the~formation of this tail is a~modification of 
the~BHL-accretion, i.e. a~gravitational focusing of the~wind that is 
basically similar to the~gravitational lensing of photons.
Unlike the~axially symmetric case of the~classical BHL-accretion, 
the~stellar wind from a~supergiant has spatial gradients of both 
the~velocity and the~density when reaching the~orbit of the~companion 
object and its accretion must thus be non-axisymmetric.  The~violation
of the~symmetry is more pronounced if the~wind speed does not 
significantly exceed the~orbital velocity of the~companion.  In this
asymmetric case, the~classical approximation of the~BHL accretion rate 
is less accurate \citep{1997ASPC..121..822W} and the~accretion is unstable
\citep{1997A&A...317..793R,1999A&A...346..861R}. 
{There are sophisticated general-relativistic radiation- or
magneto- hydrodynamic models 
\citep[cf., e.g.,][ and references therein]{2011MNRAS.417.2899Z} 
of the BHL-accretion.  These models are needed to treat in detail 
the~accretion in the close vicinity of the~compact component and 
the~generation of the~X-ray radiation.  The~present models, however, 
mostly assume the~homogeneity of the~incident flow.  An inhomogeneity 
in the~flow may significantly influence the~angular momentum of 
the~accreted matter, hence both the~size of the~inner accretion disc 
and its luminosity.  The~inhomogeneity of the~wind in the~neighbourhood 
of the~companion depends on its tidal modulation in the~subsonic region 
of the~donor star and its non-radial flow in the~space between 
the~components.  These effects have thus to be taken into account 
in self-consistent models of the~accretion.}

   The existence of these gaseous tails in interacting binaries is 
confirmed observationally -- cf. \cite{2000A&A...354.1014D} 
or \cite{2006smqw.confE..52D}.
The~deposition of the~mass focused into the~tail and
the~well-known instability of BHL accretion \citep{2005A&A...435..397F}
may influence the~interplay between the~variability of the~wind mass-loss
from the~donor star, the~hydrodynamics of the~circumstellar matter, and
its accretion onto the~compact object.  It may thus be responsible for 
different kinds of time variability for these objects
-- cf., e.g., \cite{2009ApJ...700...95B}. 

\begin{acknowledgements}
The authors are grateful to J. Dale, {K.~G.~Gayley}, J.~Kub\'{a}t, and 
R.~W\"{u}nsch for useful comments and suggestions.
This work has been done in the~framework of the~Center for Theoretical
Astrophysics (ref.~LC06014) with a~support of grants GA\v{C}R 202/09/0772
and 205/09/H033, GAUK 139810 and SVV-265301.
\end{acknowledgements}

\bibliography{AA16046}

\appendix

\section{Model atmospheres of binary components}
We now reproduce in a~language corrected form the~parts of 
Hadrava (1987) relevant to our present study.  Figure captions 
missing in the~original paper because of lack of space are added here.\\

 The~models of stellar atmospheres are based on the~assumption of 
either their plane-parallel or spherical symmetry.  The~violation 
of either of these assumptions by tides and rotation in close binaries 
leads to the~incompatibility of hydrostatic and radiative equilibria. 
An improvement in the~model atmospheres in this respect is desirable 
for simulating both the~light curves and line profile changes. Moreover, 
the~atmospheres of the~contact components of interacting binaries 
determine the~initial conditions of dynamics of gaseous streams and in 
this way influence the~behaviour of the~binary system.

\subsection{Hydrostatic equilibrium}\label{Append1}

It is well-known that the~stellar configuration in hydrostatic equilibrium
must be homogeneous across each equipotential. This follows from the~vector
nature of the~equation of the~hydrostatic equilibrium
\begin{equation}\label{Eq1}
 \rho\vec{\nabla}\Phi+{\vec \nabla} P=0\; ,
\end{equation}
because the~pressure $P$ and density $\rho$ must be functions only of 
the~potential $\Phi$, and following the~equation of state
\begin{equation}\label{Eq2}
 P=P(\rho, T)
\end{equation}
the~same must be valid for the~temperature $T=T(\Phi)$. The~temperature
distribution is determined by the~energy balance, driven mainly by 
radiative transfer.  The~solution of radiative transfer is the~main 
problem in modelling symmetric (i.e. plane-parallel or spherical) 
atmospheres, where the~tangential components of Eq.~(\ref{Eq1}) are 
trivial and both Eqs.~(\ref{Eq1}) and (\ref{Eq2}) can be simply 
integrated. However, if the~symmetry is violated e.g. by either 
rotation or tides, the~homogeneity imposes an a~priori restriction
on the~radiation field across different parts of stellar surface. 
This restriction cannot generally be satisfied.  This can be seen 
from Fig.~\ref{OO019f1}, where the~source function $S$ is plotted 
(full line) for the~grey atmosphere.  If $g$ is increased, 
the~hot inner layers are shifted to lower optical depths (dashed 
line), which in agreement with von~Zeipel's theorem corresponds 
to an~increase in the~effective temperature (dotted line).  On 
the~other hand, the~surface temperature (for $\tau\rightarrow 0$) 
remains constant and is thus incompatible with the~radiation 
coming from the~interior. When the~surface layers are heated by
absorbing this radiation, the~hydrostatic equilibrium is violated 
in the~tangential direction and tangential radiative transfer 
takes place. The~latter of these generally also appears as
a~consequence of the~different radiation dilution caused
by the~variable curvature of the equipotentials. 
% (see the next Section).
The~incompatibility of the~hydrostatic
and radiative equilibria should give rise to %a
meridional circulations in the~atmosphere.  The~construction of 
model atmospheres as a~mosaic of one-dimensional (especially 
plane-parallel) models can thus be accepted only as a~first
approximation.  Another violation of hydrostatic equilibrium 
is caused by the~existence of stellar wind, which is discussed 
in Section~\ref{Append2}. Additional complications arise when
the~binary component does not rotate synchronously with the~orbital 
motion, particularly when the~orbit is eccentric.  Despite
the~aforementioned objections, the~hydrostatic equilibrium is 
a~valuable approximation, at least for the~inner layers of 
atmospheres.  It is thus advantageous to choose a~coordinate 
system that fits the~corresponding shape of the~atmosphere for
the~further investigation. This is achieved using Roche
coordinates, where the~first coordinate $\xi_{1}$ is
the~effective potential and the~surfaces $\xi_{2,3}=const$
are perpendicular to the~equipotentials (but generally
not mutually perpendicular). The~generalisation
of a~Roche potential to a~finite mass concentration
is also possible, but fortunately not essential.

%                                                One column figure
%------------------------------------------------ Fig.1 - grav. darkening
\begin{figure}  
\centering
   \setlength{\unitlength}{1mm}
 {\includegraphics[width=89.mm]{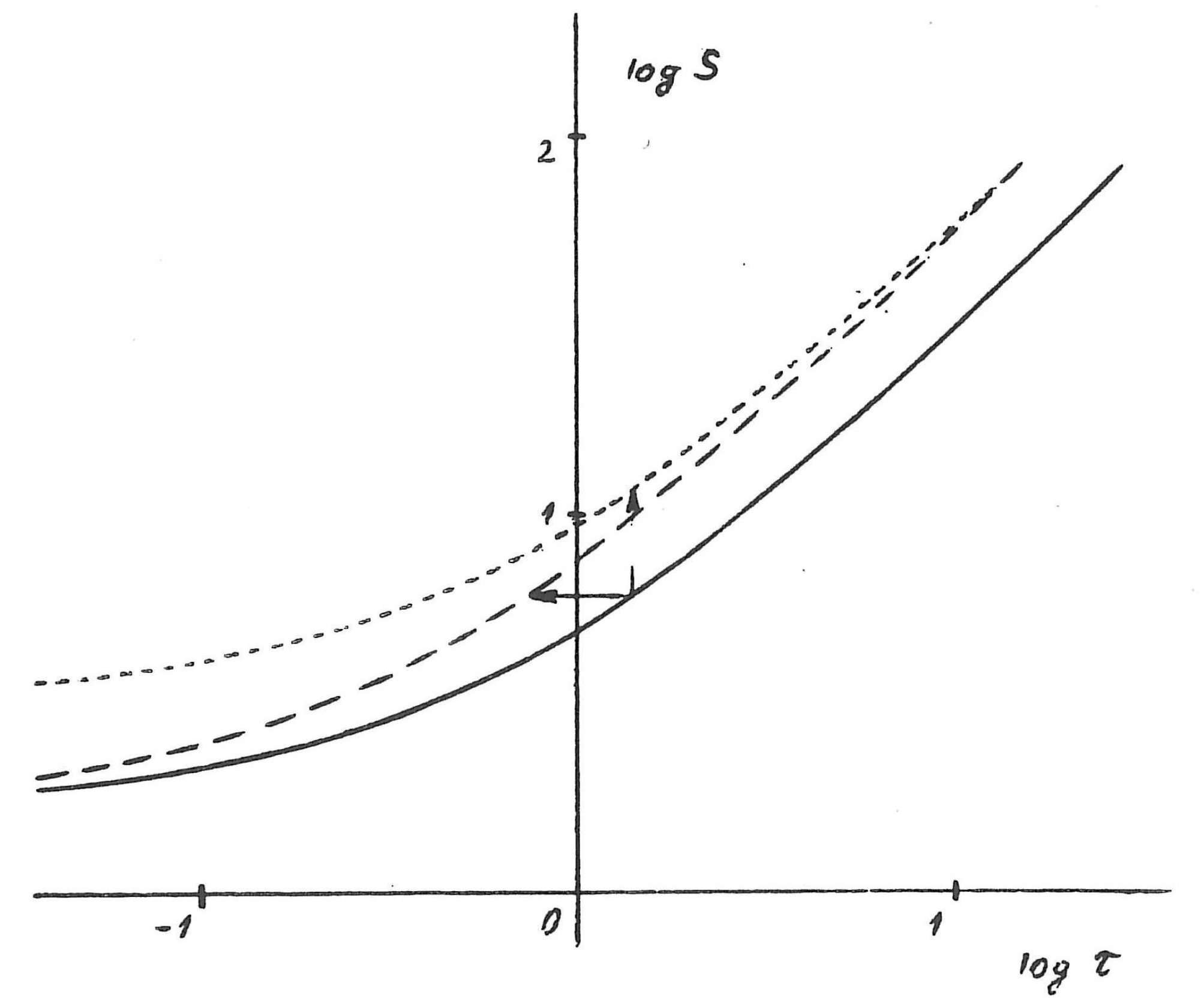}}
\caption{ %
 The~dependence of source function $S$ in a~grey atmosphere
on the~optical depth $\tau$ (the~full line). In the~parts
of atmosphere with a~higher gravity $g$, layers with a~given
temperature are shifted to lower optical depths (the~ dashed 
line). This corresponds to increase of the~radiative flux
in large optical depths (the~dotted line), which gives, however,
a~temperature incompatible at low depths with the~homogeneity
on equipotential. The~actual value of emergent flux $H$ for
different shifts of $S$ in $g$ is shown on Fig.~\ref{OO019f2}
}\label{OO019f1}
\end{figure}
%---------------------------------------------------------

%                                                One column figure
%------------------------------------------------ Fig.2 - grav. darkening
\begin{figure}

\centering
   \setlength{\unitlength}{1mm}
 {\includegraphics[width=89.mm]{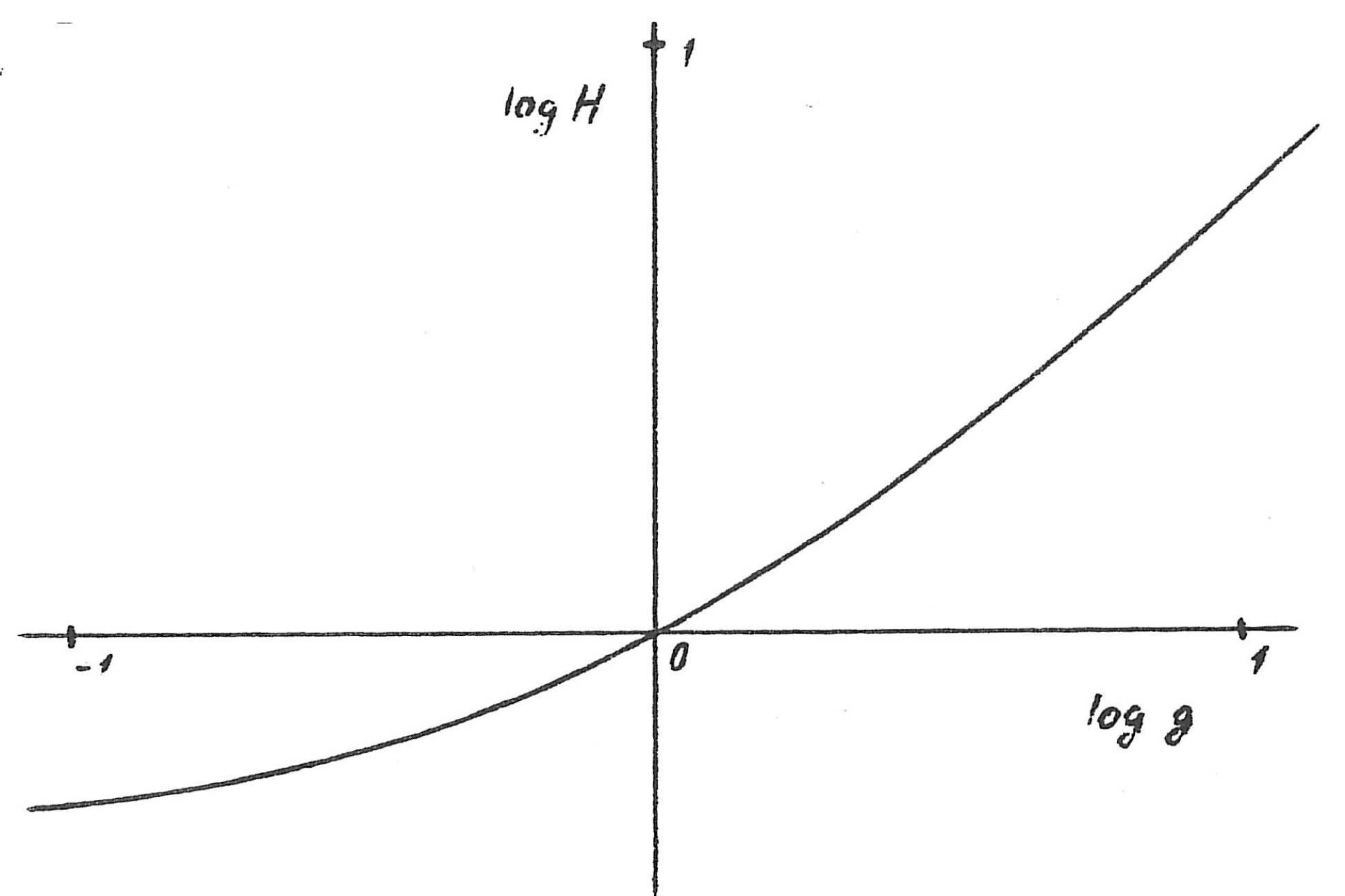}}
\caption{ %
 The~dependence of radiative flux $H$ from an~inhomogeneous
grey atmosphere in exact hydrostatic equilibrium on the~local 
gravity $g$ (in units of its mean unperturbed value).
}\label{OO019f2}
\end{figure}
%---------------------------------------------------------

\subsection{Stellar wind}\label{Append2}

The~density of a~gas in thermodynamic equilibrium is an~exponential 
function of the~potential. This is why it decreases to zero and 
the~mass (or optical) depth is finite in upper layers of plane-parallel 
atmosphere, where $\Phi\rightarrow\infty$. However, $\Phi\rightarrow0$ 
at infinity for a~spherical (or other finite) configuration and 
consequently $\rho \rightarrow\rho_{\infty}>0$ and the~mass depth 
becomes infinite for such a~configuration in equilibrium with its 
surroundings. If a~real star is immersed into vacuum, it must 
evaporate. Eq.~(\ref{Eq1}) of hydrostatic equilibrium must then be 
thus replaced by the~equation of motion
\begin{equation}\label{Eq12}
 \partial_{t}\vec{u}+(\vec{u}\vec{\nabla})\vec{u}
 +2\vec{\Omega}\wedge\vec{u}+\vec{\nabla}\Phi+\frac{1}{\rho}\vec{\nabla} P=0
\end{equation}
and the~equation of continuity
\begin{equation}\label{Eq13}
 \partial_{t}\rho+\vec{\nabla}(\rho\vec{u})=0\; .
\end{equation}
This system of partial differential equations can be simplified for 
a~stationary flow ($\partial_{t}=0$), if the~stream lines
\begin{equation}\label{Eq14}
 \vec{x}=\vec{x}(s)\; ,\hspace*{5mm}
 {\vec u}=u{\vec n}\; ,\hspace*{5mm}
 {\vec n}=\frac{\mathrm{d}\vec{x}}{\mathrm{d}s}\; ,
\end{equation}
are known (e.g. if they are radial for spherically symmetric problem).
Eq.~(\ref{Eq13}) then becomes
\begin{equation}\label{Eq15}
 \frac{\mathrm{d}}{\mathrm{d}s}(\rho u \mathcal{D}) = 0
\end{equation}
and its solution is
\begin{equation}\label{Eq16}
 \rho=\frac{h}{u\mathcal{D}}\; ,
\end{equation}
where $h$ is the~flow of gas, which is constant along the~stream line, 
and $\mathcal{D}=\mathcal{D}(s)$ is the~cross-section of stream-tube, 
determined by the~dilution of the~stream lines
\begin{equation}\label{Eq17}
 \frac{\mathrm{d}}{\mathrm{d}s}{\rm ln} \mathcal{D} =(\vec{\nabla}\vec{n})\; .
\end{equation}
The component of Eq.~(\ref{Eq12}) parallel to $n$ reads
\begin{equation}\label{Eq18}
 \frac{\mathrm{d}}{\mathrm{d}s}\frac{u^{2}}{2}+ \frac{\mathrm{d}}{\mathrm{d}s}\Phi+
 \frac{1}{\rho} \frac{\mathrm{d}}{\mathrm{d}s} P = 0\; .
\end{equation}
%                                                One column figure
%------------------------------------------------ Fig3. - radwin
\begin{figure} 
\centering
   \setlength{\unitlength}{1mm}
 {\includegraphics[width=89.mm]{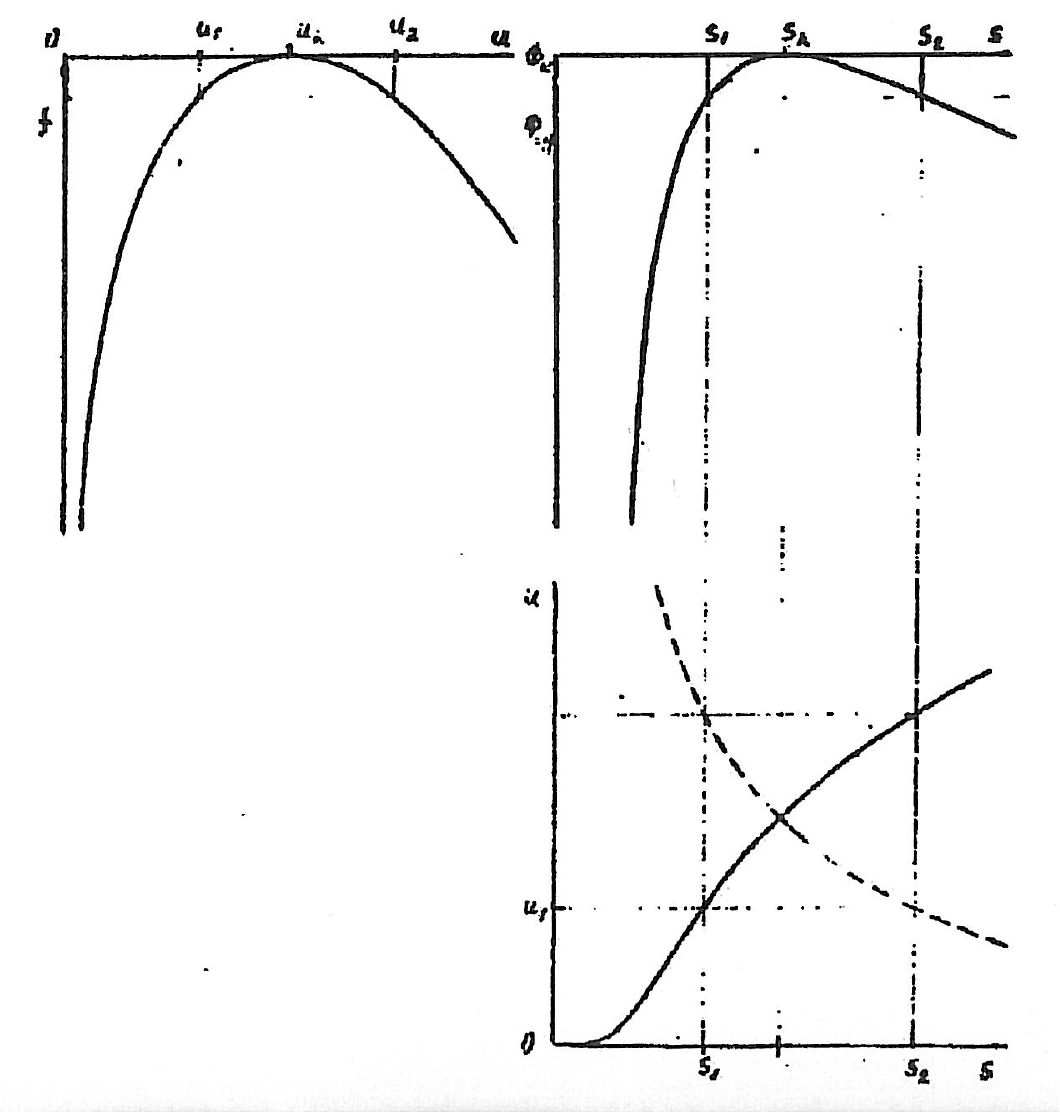}}%{OO019p02.pdf}}
\caption{ %
 Evaporative wind in an effective Roche potential -- the~left and right
 upper panels give the~left and right sides of Eq.~(\ref{Eq19}), i.e.
 the~effective potential $\Phi_{\rm eff}$ as a~function of $u$ and $s$, resp. 
 The~bottom panel gives the~solution $u=u(s)$, the~full~line corresponds
 to the~wind solution, and the~dashed line to accretion.}\label{OO019f4}
\end{figure}
%---------------------------------------------------------
The density can be substituted here from Eq.~(\ref{Eq16}), and if 
the~thermal regime of the~flow is known (e.g. where there is a~polytropic 
expansion or $T=T(s)$), the~equation for $u=u(s)$ is obtained. It can be 
simply integrated in the~form
\begin{eqnarray}\label{Eq19}
 f(u)&=&\frac{u^{2}_{k}-u^{2}}{2}+\frac{kT}{m}{\rm ln}\frac{u}{u_{k}}=\\
 &=&-\frac{kT}{m}{\rm ln} \mathcal{D}+\Phi-\Phi_{k}=\Phi_{\rm eff}(s)-\Phi_{k}\nonumber
\end{eqnarray}
for an idealised problem of isothermal flow ($\Phi_{k}$ is an integration
constant). There are two solutions $u(s)\leq u_{k}\equiv\sqrt{\frac{kT}{m}}$ 
and $u(s)\geq u_{k}$ of this equation -- see Fig.~\ref{OO019f4}. The~former is 
asymptotically hydrostatic, while the~latter corresponds to a~free fall
for $\Phi_{\rm eff}\rightarrow-\infty$. An important feature is that 
the~dilution of the~stream lines acts as a~repulsive force (nozzle) through 
the~first term on the~r.h.s. of Eq.~(\ref{Eq19}). The~above-mentioned 
solution can be matched at the~critical point $s_{k}$, where $\Phi_{\rm eff}$
reaches its maximum $\Phi_{\rm eff}(s_{k})=\Phi_{k}$, and $u(s_{k})=u_{k}$.
Its position can be influenced or even dominated by the~presence of
another repulsive force such as radiative pressure or centrifugal force. 
Thei~latter is combined with the~tidal force in binaries. A~critical point 
thus depends on the~Roche potential along a~particular stream line. It
determines the~velocity in deep layers, where the~hydrostatic equilibrium
is approached and the~density is the~same exponential function of potential
for each stream line. The~flow of mass is thus
\begin{equation}\label{Eq20}
 h\sim\exp\left(-\frac{m\Phi_{k}}{kT}\right)
\end{equation}
in different directions. The~results for a~radial wind are shown in
Fig.~\ref{OO019f5} for several values of the~mass ratio $q$ and ratio 
$\theta=
\frac{kTR}{mgM}$ of thermal kinetic to gravitational potential
energy of a~gas particle. The~Roche potential is thus a~potential
barrier to the~cool gas, that can flow only in the~vicinity of 
the~$L_{1}$-point, although the~anisotropy of the~stellar wind decreases 
with increasing surface temperature. In any case, the~Roche-lobe 
overflow and mass loss by stellar winds are not independent processes
as often assumed, but only two extreme pictures of 
the~same physical process -- an~anisotropic stellar wind.

%                                         One column figure
%------------------------------------------------ Fig.5 
\begin{figure}
\centering
   \setlength{\unitlength}{1mm}
 {\includegraphics[width=89.mm]{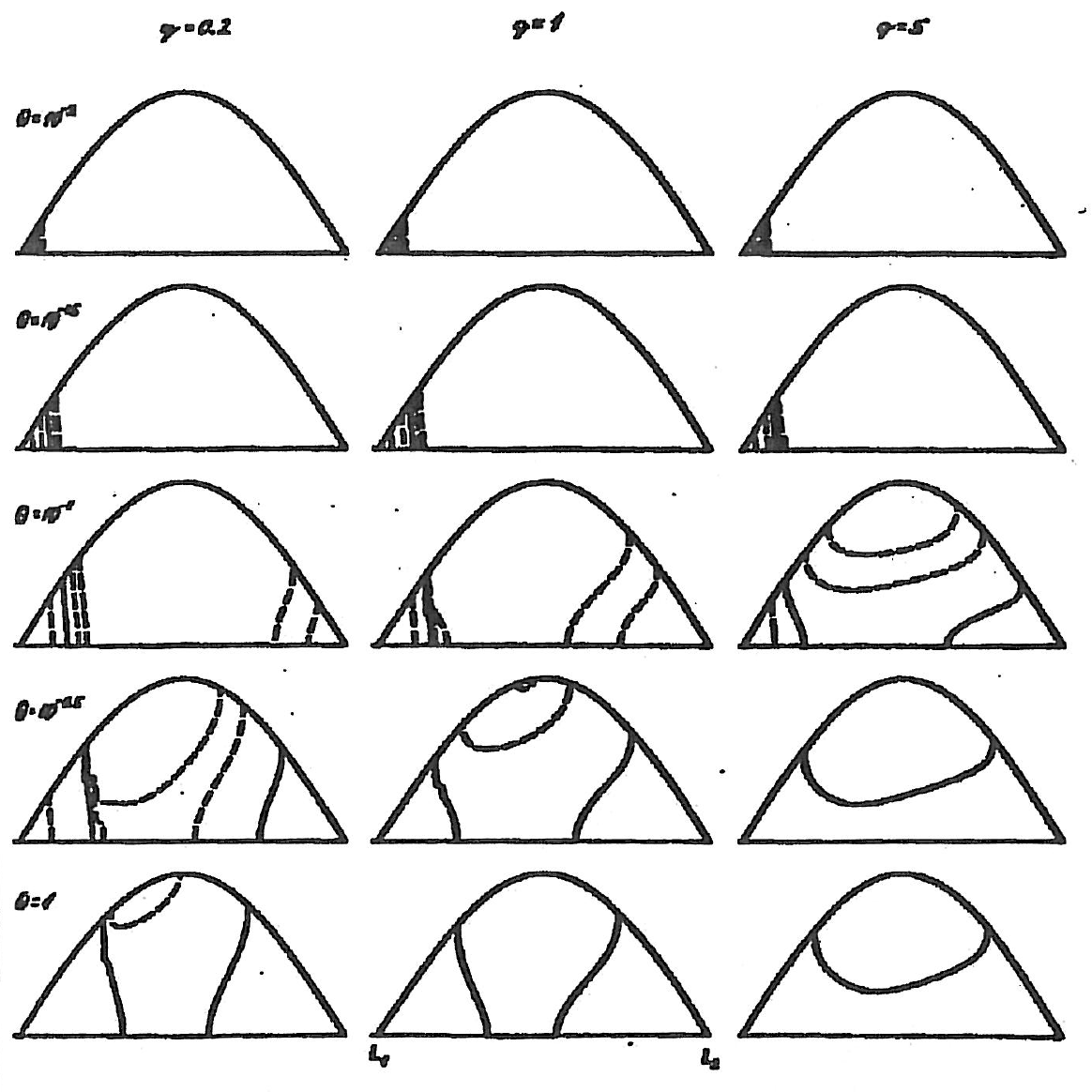}}%{OO019p01.pdf}}
\caption{ %
  Angular distribution of wind strengths for several values of 
$\theta$ and $q$.  Full lines mark the~isocontours of the~mean 
wind intensity (dashed lines show values for one and two orders 
higher and lower) in spherical coordinates with poles towards 
the~$L_{1}$ and $L_{2}$ points.  The~latitude $\varphi$ is drawn 
in a~linear scale on the~horizontal axis; the~longitude $\lambda$ 
measured from the~intersection with the~orbital plane scaled by 
$\cos \varphi$ increases upward up to the~rotational axis at 
the~top of each drawing.}\label{OO019f5}
\end{figure}
%---------------------------------------------------------

%{append}

\end{document}